\documentclass[12pt]{article}
\usepackage{amsmath}
\usepackage{amsfonts}
\usepackage{amssymb}
\usepackage{graphicx,psfrag,epsf}
\usepackage{enumerate}
\usepackage{natbib}
\usepackage{adjustbox}
\usepackage{caption}
\usepackage{subcaption}
\usepackage{multirow}
\usepackage[margin=1in]{geometry}
\usepackage[compact]{titlesec}
\usepackage{comment}
\usepackage[utf8]{inputenc}
\usepackage{listings}
\usepackage{color}

\usepackage{algorithm}
\usepackage{algpseudocode}
\usepackage{graphicx}
\usepackage{caption}
\usepackage{subcaption}

\usepackage{paralist}
\usepackage{url} 

\newcommand{\blind}{0}

\begin{document}

\def\spacingset#1{\renewcommand{\baselinestretch}%
{#1}\small\normalsize} \spacingset{1}


\if0\blind
{
  \title{\bf Two-Sample Testing for Multivariate Cross-Correlation Functions with Applications to Gut-Brain Reward Learning}
  \author{Bhaskar Ray, 
Tùng Bùi, William Matthew Howe, Srijan Sengupta}
  \maketitle
} \fi

\if1\blind
{
  \bigskip
  \bigskip
  \bigskip
  \begin{center}
    {\LARGE\bf Title}
\end{center}
  \medskip
} \fi

\begin{abstract}
Cross-correlation functions (CCFs) are widely used to explore lead-lag relationships between paired time series, but in most applications they remain primarily descriptive tools with limited formal inferential methodology. This issue is especially important in scientific settings where the dependence between two time-varying processes may appear across a broad range of positive and negative lags rather than only contemporaneously. Motivated by a mouse neuroscience study of gut-brain interactions in reward learning, we implement a two-sample inference framework for comparing collections of subject-specific CCF curves across experimental groups. In our application, each recording session yields two related CCFs describing the temporal association of dopamine activity with locomotor velocity and acceleration, respectively, leading naturally to a multivariate functional data formulation.

Our approach treats the empirical CCF as a functional object indexed by lag and uses tools from multivariate functional data analysis to test equality of mean CCF functions between groups. In particular, we employ integrated and maximum-type global test statistics, denoted by \(F_{\mathrm{int}}\) and \(F_{\max}\), based on pointwise Hotelling \(T^2\) statistics. The integrated test is sensitive to broad differences distributed over the lag domain, whereas the maximum test is more sensitive to localized departures. We apply the proposed framework to free-feeding and intragastric infusion experiments designed to study how post-ingestive nutrient signals influence dopamine dynamics in the nucleus accumbens and dorsal striatum. The analysis reveals clear differences in dopamine-locomotion coupling across brain regions and biological sex in the free-feeding experiment, with more selective effects in the infusion setting. More broadly, the paper demonstrates that viewing CCFs through a functional-data lens yields a flexible and rigorous framework for statistical comparison of dynamic dependence patterns across experimental conditions.
\end{abstract}
\bigskip

\noindent%

\spacingset{1.1} 

\maketitle

\section{Introduction}

Suppose we observe two time series, $Y_1(t)$ and $Y_2(t)$, for $t=1, \ldots, T$.
We are interested in understanding the relationship between $Y_1(t)$ and $Y_2(t)$, where one series may be related to past (lagged) or future (leading) values of the other series.
A classical tool for analyzing such relationships is the cross-correlation function or CCF.
More formally, for a bivariate weakly stationary time series \(\{(Y_1(t),Y_2(t)) : t\in \mathbb{Z}\}\), the cross-covariance function at lag \(l\) is defined as
\[
\gamma_{12}(l)=\mathrm{Cov}\{Y_1(t),Y_2(t+l)\},
\]
and the corresponding cross-correlation function (CCF) is
\[
\rho_{12}(l)=\frac{\gamma_{12}(l)}{\sqrt{\gamma_{11}(0)\gamma_{22}(0)}},
\]
where \(\gamma_{11}(0)=\mathrm{Var}\{Y_1(t)\}\) and \(\gamma_{22}(0)=\mathrm{Var}\{Y_2(t)\}\). In practice, the CCF is typically estimated empirically over a range of positive and negative lags and displayed graphically as a correlogram \citep{brockwell2002introduction,shumway2006time}. Such plots are widely used as an exploratory data analysis tool for identifying lead-lag relationships, assessing temporal association between two series, and generating hypotheses about possible mechanisms linking the underlying processes. However, in most applications the CCF is used primarily as a descriptive metric for exploratory data analysis, with comparatively less emphasis on formal inference and hypothesis testing.

In this paper, our goal is to carry out formal statistical inference for CCFs. Our motivating application comes from a mouse neuroscience study on gut--brain control of reward learning. A large literature implicates mesostriatal dopamine signaling in reinforcement learning, reward prediction, and motivated behavior, with especially prominent roles for the nucleus accumbens (NAc) and dorsal striatum (DS) \citep{schultz1998predictive,schultz2016dopamine}. More recent work suggests that post-ingestive nutrient signals arising from the gastrointestinal tract can modulate dopamine-related reward circuits and thereby influence food reward and learning \citep{dearaujo2012gut,han2018gut,goldstein2021hypothalamic}. The data analyzed here were collected as part of a broader experimental program investigating how nutrient-derived signals influence dopamine release in striatal subregions during food-related behavior. For each experimental session, dopamine activity was recorded using fiber photometry with a genetically encoded dopamine sensor while the animal's locomotor behavior was tracked simultaneously, yielding paired time series observed at high temporal resolution \citep{patriarchi2018dlight}.

This scientific setting naturally motivates the use of cross-correlation functions rather than static correlation summaries. The relationship between dopamine and locomotion is inherently dynamic and may operate in both temporal directions: changes in dopamine release may be followed by changes in future locomotion, while ongoing or impending locomotion may also be associated with subsequent dopamine fluctuations. Therefore, the statistical association between the two processes need not be concentrated at lag zero, nor even be symmetric around zero. Instead, it may manifest through leading or lagging correlations at a range of positive and negative lag values. An instantaneous correlation therefore captures only a narrow slice of the dependence structure and may miss scientifically meaningful temporal relationships. In contrast, the CCF curve summarizes association across a continuum of lead and lag values, thereby providing a more complete representation of dopamine-behavior coupling. In our application, each session gives rise to a subject-specific CCF curve over a grid of lags, and our inferential goal is to determine whether these curves differ systematically across biologically meaningful groups such as brain region, sex, and nutrient condition.

This perspective also suggests functional data analysis (FDA) as a natural methodological framework for statistical inference on CCFs. Rather than viewing the empirical CCF merely as a collection of correlations evaluated at a finite set of lags, we treat it as a realization of an underlying random function defined over a continuum of lag values. This is justified by classical large-sample theory for sample cross-correlations, since under suitable weak-dependence conditions, the empirical CCF at any fixed lag is asymptotically normal, so that pointwise the CCF behaves like a Gaussian random variable in large samples \citep{shumway2006time}. This motivates modeling the subject-specific CCF as a Gaussian process indexed by lag. Such a formulation is especially attractive because the true, population-level CCF is naturally defined for all lag values, whereas in practice we only observe noisy empirical estimates on a discrete grid. The Gaussian-process framework provides a flexible and tractable way to pass from these finitely observed values to an estimated infinite-dimensional curve, while simultaneously accommodating the covariance structure across nearby lags \citep{suzuki2022gaussian,shi2011gaussian}. In this way, FDA provides a principled mechanism for smoothing, interpolation, and statistical comparison of CCFs as functional objects rather than isolated point estimates.

The remainder of the paper is organized as follows. In Section 2, we review relevant FDA methodology and describe the statistical framework for two-sample inference with multivariate CCFs. Section 3 presents the real-data application to dopamine-locomotion recordings from the gut-brain experiment and illustrates the proposed inferential tools in practice. Section 4 concludes with a discussion of the main findings, methodological limitations, and possible directions for future work.

\section{Methodology}

We begin with a brief review of two-sample testing methodology for functional data and then describe the multivariate \(F_{\mathrm{int}}\) and \(F_{\max}\) procedures used in this paper.

\subsection{Two-sample testing in functional data analysis}

Two-sample testing is a fundamental problem in functional data analysis (FDA). Let
\[
X_1,\ldots,X_m \stackrel{\mathrm{i.i.d.}}{\sim} P,
\qquad
Y_1,\ldots,Y_n \stackrel{\mathrm{i.i.d.}}{\sim} Q,
\]
where \(X_i\) and \(Y_j\) are random functions defined on a compact interval \(\mathcal T \subset \mathbb R\), typically viewed as elements of \(L^2(\mathcal T)\). 
Then, the most general version of the two-sample test is given by
\[
H_0: P = Q
\qquad \text{versus} \qquad
H_1: P \neq Q,
\]
while a more common and more targeted formulation tests equality of the mean functions,
\[
H_0: \mu_X(t) = \mu_Y(t)\ \ \forall t \in \mathcal T,
\qquad
\mu_X(t) = \mathbb E\{X(t)\},\quad \mu_Y(t) = \mathbb E\{Y(t)\}.
\]
Unlike the classical multivariate setting, the parameter of interest is now infinite-dimensional, and testing procedures must account for smoothness, dependence across the continuum, and often discretization or smoothing of the observed curves \citep{ramsay2005fda,ramsay2009fda,kokoszka2017introduction}. 

A natural starting point is the class of mean-based global tests obtained by first constructing a pointwise statistic and then aggregating it over the domain. For example, one may form a pointwise two-sample \(t\)-statistic
\[
T(t) =
\frac{\bar X(t)-\bar Y(t)}
{\sqrt{\hat\sigma_X^2(t)/m+\hat\sigma_Y^2(t)/n}},
\qquad t\in\mathcal T,
\]
and then define a global maximum-type statistic
\[
T_{\max}=\sup_{t\in\mathcal T}|T(t)|.
\]
Such procedures are intuitive and are often calibrated by permutation, since the null distribution of \(T_{\max}\) is typically unavailable in closed form \citep{ramsay2009fda,kokoszka2017introduction}. Closely related ideas arise in functional ANOVA, where the pointwise \(F\)-statistic is globalized by taking either its supremum or an integrated version over the domain. In particular, the \(F_{\max}\) approach has been shown to perform well when within-function correlation is moderate to strong \citep{zhang2019fmax}. 

A broader class of methods addresses equality of the full distributions \(P\) and \(Q\), rather than only equality of means. These include permutation-based nearest-neighbor and depth-based procedures, which compare the geometric arrangement of curves in function space. Such methods are attractive because they can detect changes beyond simple mean shifts, including differences in variability or shape, although their performance can depend strongly on the chosen metric or depth notion \citep{cabana2017permutation}. 

There is also a substantial literature on two-sample testing for multivariate functional data, where each observational unit consists of a vector of functions observed jointly over a common domain. In that setting, the null hypothesis becomes
\[
H_0:\boldsymbol{\mu}_1(t)=\boldsymbol{\mu}_2(t)\quad \forall t\in\mathcal T,
\]
where \(\boldsymbol{\mu}_g(t)\in\mathbb R^p\) is a mean vector function. Pointwise Hotelling \(T^2\)-type statistics provide a natural multivariate analogue of the scalar-valued \(t\)-statistic, and these pointwise statistics may again be globalized either by integration over \(\mathcal T\) or by taking their supremum \citep{qiu2021multivariate}. This framework is particularly relevant in applications, such as ours, where multiple coordinated functional summaries are recorded for each experimental unit.

Finally, many classical FDA procedures assume dense observation on a common grid. This has motivated the development of methods for sparse functional data, as well as kernel-based nonparametric two-sample procedures that target distributional equality in reproducing kernel Hilbert spaces \citep{wang2021sparse,wynne2022kernel}. Overall, the FDA two-sample testing literature contains a range of tools, differing according to whether the inferential target is the mean function or the full distribution, whether the data are univariate or multivariate, and whether the curves are densely or sparsely observed \citep{ramsay2009fda,kokoszka2017introduction,cabana2017permutation,qiu2021multivariate,wang2021sparse,wynne2022kernel}. In this paper, because each session yields two related CCF curves, namely dopamine versus velocity and dopamine versus acceleration, our problem falls most naturally into the setting of two-sample testing for multivariate functional data.

\subsection{Multivariate $F_{\mathrm{int}}$ and $F_{\max}$ tests for CCF curves}

For each experimental session, let
\[
\mathbf Y_{ij}(h)=
\begin{pmatrix}
Y_{ij1}(h)\\
Y_{ij2}(h)
\end{pmatrix},
\qquad h\in\mathcal H,
\]
denote the observed bivariate CCF curve, where \(i\in\{1,2\}\) indexes the group, \(j=1,\ldots,n_i\) indexes the session within group \(i\), and \(\mathcal H=[a,b]\) denotes the lag domain of interest. In our application,
\[
Y_{ij1}(h)=\mathrm{CCF}\{\text{dopamine},\text{velocity}\}(h),\qquad
Y_{ij2}(h)=\mathrm{CCF}\{\text{dopamine},\text{acceleration}\}(h),
\]
so that \(p=2\). Thus, each session contributes a vector-valued function rather than a scalar-valued one, and the relevant inferential target is the mean vector CCF function
\[
\boldsymbol{\mu}_i(h)=\mathbb E\{\mathbf Y_{ij}(h)\}\in\mathbb R^2.
\]
The two-sample null hypothesis is
\[
H_0:\boldsymbol{\mu}_1(h)=\boldsymbol{\mu}_2(h)\quad \forall h\in\mathcal H,
\]
against
\[
H_1:\boldsymbol{\mu}_1(h)\neq \boldsymbol{\mu}_2(h)\quad \text{for some } h\in\mathcal H.
\]
This formulation allows us to test whether the joint dopamine-velocity and dopamine-acceleration cross-correlation structure differs across experimental conditions.
Following \citet{qiu2021multivariate}, we assume that within each group the multivariate functional observations are independent realizations from a \(p\)-dimensional stochastic process,
\[
\mathbf Y_{i1}(h),\ldots,\mathbf Y_{in_i}(h)\stackrel{\mathrm{i.i.d.}}{\sim} SP_p(\boldsymbol{\mu}_i,\Gamma),
\qquad i=1,2,
\]
where \(\Gamma(s,t)\in\mathbb R^{p\times p}\) is a common covariance matrix function. In our setting \(p=2\), but we present the methodology for general \(p\).

Next, we describe the derivation of the pointwise Hotelling \(T^2\) statistic.
For each lag value \(h\in\mathcal H\), let
\[
\bar{\mathbf Y}_{i\cdot}(h)=\frac{1}{n_i}\sum_{j=1}^{n_i}\mathbf Y_{ij}(h),
\qquad i=1,2,
\]
denote the group mean vector functions, and let
\[
\widehat{\Gamma}(s,t)
=
\frac{1}{n-2}\sum_{i=1}^2\sum_{j=1}^{n_i}
\bigl(\mathbf Y_{ij}(s)-\bar{\mathbf Y}_{i\cdot}(s)\bigr)
\bigl(\mathbf Y_{ij}(t)-\bar{\mathbf Y}_{i\cdot}(t)\bigr)^\top,
\]
where \(n=n_1+n_2\), denote the pooled covariance matrix function estimator. Then the pointwise Hotelling statistic is
\[
T_n(h)
=
\frac{n_1n_2}{n}
\bigl(\bar{\mathbf Y}_{1\cdot}(h)-\bar{\mathbf Y}_{2\cdot}(h)\bigr)^\top
\widehat{\Gamma}(h,h)^{-1}
\bigl(\bar{\mathbf Y}_{1\cdot}(h)-\bar{\mathbf Y}_{2\cdot}(h)\bigr).
\]
This is the multivariate analogue of the pointwise squared \(t\)-statistic and measures separation between the two mean vector functions at lag \(h\), while accounting for the within-lag covariance between the two CCF components \citep{qiu2021multivariate}.

A pointwise Hotelling test at each lag would be cumbersome and would not directly answer the global question of whether the two mean vector CCFs are equal over the entire lag domain. To obtain global tests, we aggregate \(T_n(h)\) over \(h\) in two ways:
\[
F_{\mathrm{int}} \equiv T_n^{(I)} = \int_{\mathcal H} T_n(h)\,dh,
\qquad
F_{\max} \equiv T_n^{(\max)} = \sup_{h\in\mathcal H} T_n(h).
\]
The integrated statistic \(F_{\mathrm{int}}\) is most sensitive to broad, distributed differences across the lag domain, while the maximum statistic \(F_{\max}\) is most sensitive to localized departures concentrated near a smaller set of lag values \citep{qiu2021multivariate,zhang2019fmax}. In practice, when the CCF is observed on a discrete lag grid \(h_1,\ldots,h_M\), the integral is approximated numerically, for example,
$
F_{\mathrm{int}}
\approx
\sum_{m=1}^M T_n(h_m)\,\Delta h,
$
and
$
F_{\max}=\max_{1\le m\le M} T_n(h_m)
$.

Under regularity conditions and the null hypothesis, \citet{qiu2021multivariate} show that
\[
T_n^{(I)} \overset{d}{\longrightarrow} T^*
=
\int_{\mathcal H}\|\mathbf w(h)\|^2\,dh,
\qquad
T_n^{(\max)} \overset{d}{\longrightarrow} T^*_{\max}
=
\sup_{h\in\mathcal H}\|\mathbf w(h)\|^2,
\]
where \(\mathbf w(h)\) is a mean-zero Gaussian process with covariance function
\[
\Gamma^*(s,t)=\Gamma(s,s)^{-1/2}\Gamma(s,t)\Gamma(t,t)^{-1/2}.
\]
Moreover, the asymptotic distribution of \(T_n^{(I)}\) is a chi-square-type mixture,
\[
T^* \overset{d}{=} \sum_{r=1}^\infty \lambda_r A_r,
\qquad A_r \stackrel{\mathrm{i.i.d.}}{\sim} \chi^2_1,
\]
where \(\lambda_r\) are the eigenvalues of \(\Gamma^*\) \citep{qiu2021multivariate}. This motivates the Welch--Satterthwaite \(\chi^2\) approximation for calibrating \(F_{\mathrm{int}}\). Specifically, one approximates the null law of \(F_{\mathrm{int}}\) by that of \(\beta\chi^2_d\), where \(\beta\) and \(d\) are chosen to match the asymptotic mean and variance. The required quantities are then estimated from the pooled sample covariance function \citep{qiu2021multivariate,zhang2013hypothesis,zhang2014oneway}.

For \(F_{\max}\), the null distribution is more difficult to approximate analytically. Accordingly, we use the nonparametric bootstrap approach of \citet{qiu2021multivariate}, building on the bootstrap strategy for functional \(F_{\max}\) tests in \citet{zhang2019fmax}. Let
\[
\widehat{\mathbf v}_{ij}(h)=\mathbf Y_{ij}(h)-\bar{\mathbf Y}_{i\cdot}(h)
\]
denote the estimated subject-effect functions. Bootstrap samples are generated by resampling these centered subject-effect functions within groups, recomputing the bootstrap test statistic each time, and estimating the null distribution from the empirical distribution of the resulting bootstrap replicates. In our implementation, \(F_{\max}\) is calibrated using 1000 bootstrap replicates.

In our application, the scientific question is not only whether dopamine-velocity CCFs differ across groups or whether dopamine-acceleration CCFs differ across groups when considered separately, but whether the \emph{joint} temporal coupling structure differs. Because velocity and acceleration are related but distinct metrics of locomotion behavior, analyzing them jointly can capture group differences that may be distributed across the two components. The multivariate FDA framework is designed precisely for this situation, where the data from each session is represented as a vector of curves observed over the same lag domain, and the null hypothesis concerns equality of the corresponding mean vector functions \citep{qiu2021multivariate}. 
In our setting, the lag domain is a symmetric interval around zero, and the CCFs are evaluated on a grid of lead and lag values. Thus, for each session, the pair of estimated CCFs forms a bivariate functional observation on \(\mathcal H\). The methods described above then provide formal global tests for whether the mean multivariate CCF differs across two groups such as brain regions, female versus male within a region, etc.

\section{Application to gut-brain dopamine signals}

We applied the proposed methodology to CCFs of dopamine-locomotion time series obtained from mouse experiments designed to study gut-brain contributions to reward learning. The broader scientific objective of the experimental program is to understand how post-ingestive nutrient signals influence dopamine dynamics in striatal subregions that are central to reward-related behavior, especially the nucleus accumbens (NAc) and dorsal striatum (DS) \citep{schultz2016dopamine,dearaujo2012gut,han2018gut}. From a statistical perspective, each recording session produces a bivariate time series in which one component is a dopamine-related fluorescence signal and the other is a locomotor measure derived from video-based positional tracking. The observational unit in our analysis is therefore a session-specific curve-valued summary of temporal association between these two processes.

The analyses reported here use two related datasets. The first is a \emph{free-feeding} dataset, in which the mice were given food at the beginning of each session and the final 19 minutes of each session were analyzed. The second is an \emph{intragastric (IG) infusion} dataset, in which the 20 minutes immediately following nutrient infusion were analyzed. In both settings, dopamine activity was paired with locomotor summaries computed from the simultaneously recorded behavioral trajectory. Our primary locomotion variables were velocity and signed acceleration, where acceleration was obtained as the temporal derivative, or first discrete difference, of velocity. Because signed acceleration tends to average near zero over long windows due to cancellation between acceleration and deceleration phases, some exploratory analyses also considered absolute acceleration. However, the main multivariate analyses in this paper focus on velocity together with signed acceleration. For each session and each locomotion measure, we computed the empirical cross-correlation function between locomotion and dopamine over a lag window of \([-1,1]\) seconds, represented on a grid of 41 lag values. Thus, each session contributes either one CCF curve in univariate analyses or a pair of CCF curves in multivariate analyses.

Each session is also associated with several categorical covariates. The primary grouping variables are recording region (NAc versus DS), biological sex, and nutrient condition. In the free-feeding data, nutrient conditions correspond to macronutrient-based food categories such as carbohydrate, fat, and combined fat-plus-carbohydrate foods. In the IG infusion data, the conditions include nutrient infusions and a control condition (PBS), with some labels corresponding to carbohydrate/sucrose and fat/lipid conditions. These groupings define the two-sample and subgroup comparisons of interest. Our inferential target is not a scalar summary such as peak correlation or correlation at a prespecified lag, but rather the entire session-specific CCF viewed as a functional object. To compare groups, we used two complementary global tests: the integrated test \(F_{\mathrm{int}}\), which is sensitive to broad differences across the lag domain, and the maximum test \(F_{\max}\), which is sensitive to more localized departures. 
For the multivariate analyses, we jointly analyzed the velocity- and signed-acceleration-based CCFs, while univariate analyses were also carried out for each locomotion measure separately.

\subsection{Free-feeding experiment}

We begin with the free-feeding dataset. The strongest signal in this experiment came from the comparison between brain regions, namely NAc versus DS. Table~\ref{tab:freefeeding_full} summarizes the full-curve testing results. In the multivariate analysis using the velocity- and signed-acceleration-based CCFs together, both \(F_{\mathrm{int}}\) and \(F_{\max}\) were highly significant, providing strong evidence that the dopamine-locomotion association differs between NAc and DS. This conclusion was unchanged in the univariate analyses: both the velocity-based CCF and the signed-acceleration-based CCF showed significant differences across brain regions. Thus, the regional effect is not confined to a single locomotion metric, but is present in both components of the CCF analysis.

We next examined biological sex within each brain region. Within NAc, the joint bivariate analysis was significant under both \(F_{\mathrm{int}}\) and \(F_{\max}\). However, the univariate results reveal that this difference is driven mainly by the signed-acceleration CCF, since the acceleration-based analysis was highly significant under both tests, whereas the velocity-based analysis was not significant at the 5\% level. Within DS, by contrast, sex differences were significant in the joint analysis and remained significant in both univariate analyses, although the evidence was again much stronger for signed acceleration than for velocity. Taken together, these results indicate that sex-related differences in dopamine-locomotion coupling are present in both brain regions, but are especially pronounced for the acceleration-based CCF.

\begin{table}[ht]
\centering
\caption{Free-feeding data: significance results for full CCF curves. Here \(F_{\mathrm{int}}\) denotes the integrated test and \(F_{\max}\) denotes the maximum test. The bivariate analysis uses the joint CCFs for velocity and signed acceleration.}
\label{tab:freefeeding_full}
\begin{tabular}{llcccc}
\hline
Comparison & CCF(s) used & \(F_{\mathrm{int}}\) & \(p\)-value & \(F_{\max}\) & \(p\)-value \\
\hline
Brain region (NAc vs DS) & Velocity + signed acceleration & 23.75 & \(<0.001\) & 69.12 & \(<0.001\) \\
Brain region (NAc vs DS) & Velocity only & 6.88 & 0.004 & 16.92 & \(<0.001\) \\
Brain region (NAc vs DS) & Signed acceleration only & 13.82 & \(<0.001\) & 69.11 & \(<0.001\) \\
\hline
Sex within NAc (F vs M) & Velocity + signed acceleration & 10.76 & 0.002 & 28.01 & \(<0.001\) \\
Sex within NAc (F vs M) & Velocity only & 2.96 & 0.072 & 6.39 & 0.057 \\
Sex within NAc (F vs M) & Signed acceleration only & 9.78 & \(<0.001\) & 26.48 & \(<0.001\) \\
\hline
Sex within DS (F vs M) & Velocity + signed acceleration & 12.96 & \(<0.001\) & 85.98 & \(<0.001\) \\
Sex within DS (F vs M) & Velocity only & 3.68 & 0.047 & 8.82 & 0.015 \\
Sex within DS (F vs M) & Signed acceleration only & 10.21 & \(<0.001\) & 75.28 & \(<0.001\) \\
\hline
\end{tabular}
\end{table}

Overall, Table~\ref{tab:freefeeding_full} shows that the free-feeding experiment yields clear and consistent evidence of group differences in entire CCF curves. The brain-region comparison is highly significant across all specifications, and the sex comparisons are also compelling, particularly when signed acceleration is included. These findings suggest that the temporal relationship between dopamine activity and locomotion differs both by anatomical location and by sex, with acceleration appearing to be a more sensitive marker of these differences than velocity alone.

To localize where these differences arise, we also examined lag-segment analyses of the free-feeding CCFs. For the brain-region comparison, the bivariate tests were significant across all lag windows considered, indicating that the NAc--DS difference is distributed broadly over the lag domain rather than being confined to a single segment. Similar segmented analyses for sex within brain region showed that the corresponding differences were also spread across multiple lag intervals, though not always uniformly. In particular, the signed-acceleration CCF displayed stronger and more persistent evidence of group differences across windows than the velocity-based CCF. These segmented findings reinforce the conclusion that acceleration contributes more strongly than velocity to the observed group differences. By contrast, food-type comparisons in the free-feeding data were more mixed and less stable across specifications. Accordingly, the clearest and most reproducible findings in the free-feeding experiment are the strong brain-region effect and the within-region sex effect, especially when signed acceleration is included.

\subsection{IG infusion experiment}

We next report results from the IG infusion data in Table~\ref{tab:ig_modified}. 
The first main result is that the brain-region comparison became significant under both \(F_{\mathrm{int}}\) and \(F_{\max}\), indicating that the velocity-based dopamine-locomotion CCF differs between NAc and DS in the infusion setting as well. In contrast, the overall sex effect was weaker: \(F_{\mathrm{int}}\) was only borderline significant, while \(F_{\max}\) did not reject. Thus, unlike the free-feeding experiment, the dominant signal in the modified IG infusion analysis is the regional effect rather than a robust sex effect.

The food-type analyses revealed a more selective pattern. When each food type was compared against the remaining sessions, the lipid condition showed significance under both \(F_{\mathrm{int}}\) and \(F_{\max}\), while sucrose was significant only under \(F_{\mathrm{int}}\). After stratifying by brain region, the most consistent NAc signals arose for lipid and sucrose, whereas in DS the clearest effect involved the combo condition. In particular, lipid versus rest within NAc was significant under both tests, sucrose versus rest within NAc was also significant under both tests, and combo versus rest within DS was significant under both tests. These subgroup findings suggest that food-related differences in CCF structure may be present, but that they depend strongly on brain region and are not uniform across all conditions.

\begin{table}[ht]
\centering
\caption{IG infusion data: significance results for velocity-based CCF curves after interpolation of position values.}
\label{tab:ig_modified}
\begin{tabular}{lcccc}
\hline
Comparison & \(F_{\mathrm{int}}\) & \(p\)-value & \(F_{\max}\) & \(p\)-value \\
\hline
Brain region (NAc vs DS) & 2.806 & 0.032 & 19.39 & \(<0.001\) \\
Biological sex (F vs M) & 2.497 & 0.05+ & 7.61 & 0.145 \\
\hline
Food type: Lipid vs rest & 3.107 & 0.020 & 12.319 & 0.017 \\
Food type: Sucrose vs rest & 2.539 & 0.047 & 8.134 & 0.129 \\
Food type: Combo vs rest & 1.692 & 0.159 & 6.428 & 0.242 \\
Food type: PBS vs rest & 1.353 & 0.252 & 4.881 & 0.434 \\
\hline
Food type within NAc: Lipid vs rest & 6.717 & 0.0004 & 18.461 & \(<0.001\) \\
Food type within NAc: Sucrose vs rest & 3.021 & 0.037 & 11.62 & 0.020 \\
Food type within DS: Combo vs rest & 4.692 & \(2.6\times 10^{-5}\) & 13.994 & 0.020 \\
\hline
Sex within NAc (F vs M) & 2.847 & 0.045 & 9.582 & 0.050 \\
Sex within DS (F vs M) & 0.701 & 0.665 & 10.056 & 0.097 \\
\hline
\end{tabular}
\end{table}

At the same time, the IG infusion results are less uniform than those from the free-feeding experiment. Several comparisons were significant under one of the two tests but not the other.
Overall, these results provide suggestive evidence that the proposed FDA-based procedures can detect meaningful differences in CCF structure, but they do not have the same level of stability as the free-feeding results.

\subsection{Summary of empirical findings}

Across the two experiments, the proposed testing framework successfully identified differences in entire CCF curves across biologically meaningful groups. The clearest and most reproducible findings came from the free-feeding dataset, where both brain region and biological sex showed strong effects, especially when the signed-acceleration CCF was included. The modified IG infusion analysis also revealed a significant brain-region effect and some selective food-type differences, although those results were more heterogeneous. Taken together, these applications illustrate the value of combining \(F_{\mathrm{int}}\) and \(F_{\max}\): the integrated test provides sensitivity to broad, distributed differences across the lag domain, while the maximum test is better suited to detecting sharper local deviations. In this sense, the two procedures offer complementary views of how dopamine-locomotion coupling varies across experimental factors and biological features.

\section{Discussion}

In this paper, we carried out a formal two-sample inference framework for cross-correlation functions by combining ideas from time series analysis and functional data analysis. Our central idea is to treat each subject-specific CCF as a functional object rather than as a collection of isolated point estimates or a single summary statistic. This allows one to compare temporal dependence structures between paired time series over an entire range of leads and lags. Methodologically, we formulated the problem as a two-sample test for multivariate functional data, which is especially appropriate in our application because each experimental session yields two related CCF curves: dopamine versus velocity and dopamine versus acceleration. We then adopted the multivariate \(F_{\mathrm{int}}\) and \(F_{\max}\) procedures to test equality of the mean vector CCF functions across groups. In the gut-brain application, these methods identified clear and biologically interpretable differences in dopamine-locomotion coupling, particularly across brain region and biological sex in the free-feeding experiment, and more selectively in the IG infusion study. 

There are several directions for future research. First, the current work focuses on two-sample testing of mean multivariate CCF functions, but the same general framework could be extended to multi-group comparisons, covariate-adjusted models, or hierarchical settings that explicitly account for repeated sessions within animal. Second, while the present methodology treats the empirical CCF curves as observed functional responses, it would be valuable to develop theory that more explicitly propagates uncertainty from the first-stage estimation of the CCF itself, especially when the underlying time series are short, noisy, or partially observed. Third, the methods here target global differences in mean CCFs; future work could investigate simultaneous confidence bands, localized post hoc procedures, or feature-specific inference for scientifically meaningful quantities such as peak lag, sign changes, or asymmetry between leading and lagging dependence. Finally, from an applied perspective, the gut-brain experiments considered here suggest broader opportunities for FDA-based inference on functional summaries derived from neuroscience time series, including settings with additional behavioral channels, multiregion recordings, or more complex temporal structure. These directions would further strengthen the role of FDA as a general inferential framework for dynamic cross-system relationships.

\bibliographystyle{apalike}
\bibliography{ref}

\end{document}